\documentclass[11pt]{article}

\usepackage{color,soul}
\usepackage{ulem}
\usepackage{color}
\begin{document}
\begin{center}
\large \textbf{The correspondence between shadow and test field in a four-dimensional charged Einstein-Gauss-Bonnet black hole}
\end{center}

\begin{center}
Deyou Chen $^{a}$\footnote{E-mail: {deyouchen@hotmail.com}},
Chuanhong Gao $^{a}$\footnote{E-mail: {chuanhonggao@hotmail.com}},
Xianming Liu $^{b,c}$\footnote{E-mail: {xianmingliu@mail.bnu.edu.cn}},
Chengye Yu $^{a}$\footnote{E-mail: {chengyeyu1@hotmail.com}}
\end{center}

\begin{center}
$^{a}$ School of Science, Xihua University, Chengdu 610039, China\\
$^{b}$ Key Laboratory of Quark and Lepton Physics (MOE) and Institute of Particle Physics, Central China Normal University, Wuhan 430079, China\\
$^{c}$ Department of Physics, Hubei Minzu University, Enshi 445000, China\\
\end{center}

{\bf Abstract:} In this paper, we investigate the photon sphere, shadow radius and quasinormal modes of a four-dimensional charged Einstein-Gauss-Bonnet black hole. The perturbation of a massless scalar field in the black hole's background is adopted. The quasinormal modes are gotten by the $6th$ order WKB approximation approach and shadow radius, respectively. When the value of the Gauss-Bonnet coupling constant increase, the values of the real parts of the quasinormal modes increase and those of the imaginary parts decrease. The coincidence degrees of quasinormal modes derived by the two approaches increases with the increase of the values of the Gauss-Bonnet coupling constant and multiple number. It shows the correspondence between the shadow and test field in the four-dimensional Einstein-Gauss-Bonnet-Maxwell gravity. The radii of the photon sphere and shadow increase with the decrease of the Gauss-Bonnet coupling constant.

\section{Introduction}

Shadows are an important feature of black holes. The research and capture of the shadows are helpful for us to understand the fundamental properties of black holes. Recently, the first image of a black hole in the center of galaxy $M87$ has been captured by the Event Horizon Telescope Collaboration \cite{EHT}. It is a dark region in a bright background, which is formed by the black hole's accretion of photons and strong gravitational lensing effect. This work provides a strong evidence for the existence of black holes in our universe. In the aspect of theoretical research, the expressions of shadow radii of some black holes have been gotten and the influence of the black holes' parameters on the shapes and sizes of shadows has been found.

Quasinormal modes (QNMs), which characterize discontinuous complex frequencies, are asymptotic solutions of perturbation fields around compact objects under certain boundary conditions. Their real parts maintain oscillations and the imaginary parts determines oscillation feature (damped oscillation or undamped oscillation). As an important feature of black holes dynamics, they attract people's attention. One of the reasons is that the dominant QNMs can be seen in the gravitational wave signals from black holes. Through the observation of the QNMs by LIGO/Virgo laboratories, astronomers and physicists found the gravitational wave signals, which indirectly verified the existence of the black holes \cite{BPA}. Another reason involves the AdS/CFT correspondence \cite{JMM,HH1,HH2,KSS,MM}. It shows that QNMs of a $(D+1)$-dimensional asymptotically anti-de Sitter spacetime are poles of the retarded Green's function in the $D$-dimensional dual conformal field theory. This correspondence has been successfully applied to the research of various properties of strongly coupled quark-gluon plasmas. In addition, QNMs also play an important role in the quantization of areas of black holes. Using Bohr's correspondence principle in the ringing frequencies, Hod quantized the area of the Schwarzschild black hole by the relationship between the real part of the QNMs and the black hole's mass and obtained the area spacing as $4 l^2_p ln3$ \cite{SHOD1}. Subsequently, Dreyer and Kunstatter quantized the Bekenstein-Hawking entropies by the loop quantum gravity and semi-classical method \cite{SHOD2,SHOD3}, respectively. Considering the transformation between arbitrary states of a black hole, Maggiore regarded the black hole as a damped harmonic oscillator and found that its area spectrum is related to the physical frequency of the harmonic oscillator \cite{SHOD4}.

There are various approaches to derive QNMs\footnote{The detailed discussion on this aspect is referred to \cite{KSZ1} and the references therein.}. Different approaches produce different preciseness \cite{BBM,SW,EWL,IW,KS1,KS2,KS3,HPN1,HPN2,IW1,BKK9,CCDN1,BKK6,BKK2,BKK3,CJ,IW2,BKK1,BKK4,BKK5,LLQ,BKK8,MBS11,KSZ2}. Null geodesics are useful tools to obtain QNMs. The angular velocities at the unstable null geodesics orbits of the black holes determine the real parts of the QNMs. The second derivatives of the effective potentials for radial motions are introduced to express the Lyapunov exponents in the imaginary parts. In the eikonal limit, the relation between the QNMs and null geodesics was gotten by Cardoso et al. \cite{CMBWZ}. However, the correspondence between QNMs in the eikonal limit and null geodesics does not always exist. As shown in \cite{Konoplya1,Konoplya2}, this correspondence is guaranteed only for the test fields, while for the gravitational and other non-minimally coupled fields it may not be fulfilled. Shadow radii are closely related to QNMs. Through the relationship between the photon spheres and shadow radii, Jusufi expressed the real parts of the QNMs in the eikonal limit by the shadow radii \cite{JAAM1,JAAM2}. This shows a correspondence between the shadows and test fields in the spacetimes of the black holes. And then, he studied the influence of the perfect fluid dark matter parameter $k$ on the QNMs by the massless scalar and electromagnetic field perturbations, and found the value of the reflecting point $k_0$ corresponding to the maximal values for the real parts. This work is helpful to detect indirectly the dark matter near the event horizon. Subsequently, this correspondence was verified by the $3rd$ order WKB approximation approach in \cite{JAAM3,JAAM4} and by the $13th$ order WKB approximation approach in \cite{GM}, respectively.

In this paper, we investigate the photon sphere, shadow radius and QNMs of a four-dimensional charged Einstein-Gauss-Bonnet black hole. The QNMs are calculated by the $6th$ order WKB approximation approach and shadow radius, respectively. The influence of the Gauss-Bonnet coupling constant on the shadow radius and QNMs is discussed. Einstein-Gauss-Bonnet theory is a simple generalization of the Hilbert Lagrangian for high-dimensional spacetime. It was obtained by adding the Bach-Lanczos Lagrangian to the Hilbert Lagrangian. In Einstein-Gauss-Bonnet gravity, the higher-dimensional black hole solutions have been gotten. Four-dimensional black hole solutions have been a challenge until recently when the work of \cite{GL} was presented. To get a four-dimensional black hole solution, Glavan and Lin rescaled the coupling constant of the Gauss-Bonnet term, $\alpha \to \frac{\alpha}{D-4}$, and let $D \to 4$. Then they searched for the spherically symmetric vacuum solution. This four-dimensional Einstein-Gauss-Bonnet gravity theory was further improved and a consistent theory was proposed in \cite{AGM}. The generalization to other black hole solutions are referred to \cite{PGS,WL1,WL2,WL3,WL4,WL6,WL8,WL9,WL10,WL11,WL12,WL13,WL14,HosseiniMansoori:2020yfj,WL15,WL16}.

The rest is organized as follows. In the next section, we investigate the photon sphere and shadow radius of the four-dimensional charged Einstein-Gauss-Bonnet black hole. In Section 3, we calculate the QNMs by the $6th$ order WKB approximation approach and shadow radius, respectively. The effect of the Gauss-Bonnet coupling constant is discussed. The last section is devoted to our conclusions.

\section{The photon sphere and shadow radius of the charged Einstein-Gauss-Bonnet black hole}

The action for the Einstein-Gauss-Bonnet gravity with electromagnetic field in D-dimensional spacetime is described by

\begin{eqnarray}
\mathcal{S} &=& \frac{1}{16\pi}\int{dx^{D}\sqrt{-g}\left[R -F_{\mu\nu}F^{\mu\nu}+\alpha\left(R^2 -4R_{\mu\nu}R^{\mu\nu}+R_{\mu\nu\beta\gamma}R^{\mu\nu\beta\gamma}\right) \right]},
\label{eq2.1}
\end{eqnarray}

\noindent where $\alpha$ is the Gauss-Bonnet coupling constant, and $F_{\mu\nu} = \partial_{\mu}A_{\nu}-\partial_{\nu}A_{\mu}$ is the Maxwell tensor. The spherically symmetric vacuum solution in four-dimensional spacetime was first gotten by Glavan and Lin. Subsequently, various solutions of black holes in the four-dimensional Einstein-Gauss-Bonnet gravity were obtained. The charged Einstein-Gauss-Bonnet black hole in the four-dimensional spacetime is given by

\begin{eqnarray}
ds^2 = -f(r)dt^2 + \frac{1}{f(r)}dr^2 + r^2 (d\theta^2+\sin^2\theta d\phi^2),
\label{eq2.2}
\end{eqnarray}

\noindent with the gauge potential $A_{\mu}=-\frac{Q}{r}dt$, where

\begin{eqnarray}
f(r)= 1+\frac{r^2}{2\alpha}\left[1\pm \sqrt{1+4\alpha\left(\frac{2M}{r^3}-\frac{Q^2}{r^4}\right)} \right],
\label{eq2.3}
\end{eqnarray}

\noindent $M$ and $Q$ are the mass and charge of the black hole, respectively. In the above equation, "$+/-$" are labeled as the "positive/negative" branches. When it is in the far region and the coupling constant $\alpha$ disappears, the Reissner-Nordstr$\ddot{o}$m black hole is recovered by the negative branch. Therefore, we focus our attention on the negative branch. From $f(r)=0$, the event horizon ($r_+$) and inner horizon ($r_-$) are gotten as follows

\begin{eqnarray}
r_{\pm}=M\pm \sqrt{M^2-Q^2-\alpha}.
\label{eq2.4}
\end{eqnarray}

\noindent Here, $\alpha$, $M$ and $Q$ must obey certain conditions to avoid the disappearance of the event horizon. For convenience, we let $M = 1$ in this paper. When $0 < Q < \sqrt{3/2}$, $\alpha$ obeys $Q^2-4-2\sqrt{4-2Q^2}< \alpha < 1-Q^2$, and there are two horizons $r_{\pm}$. When $\sqrt{3/2} < Q < \sqrt{2}$, $\alpha$ satisfies $Q^2-4-2\sqrt{4-2Q^2}< \alpha < Q^2-4+2\sqrt{4-2Q^2}$. In the region where $\sqrt{3/2} < Q < \sqrt{2}$ and $Q^2-4-2\sqrt{4-2Q^2}< \alpha < Q^2-4+2\sqrt{4-2Q^2}$, there is only one horizon $r_{+}$.

We first investigate a free photon orbiting around the equatorial orbit of the black hole (\ref{eq2.2}). The photon moves along a null geodesic. Its Lagrangian is

\begin{eqnarray}
\mathcal{L} &=& \frac{1}{2}\left[-f(r)\dot{t}^2 + \frac{1}{f(r)}\dot{r}^2 + r^2\dot{\phi}^2\right].
\label{eq3.1}
\end{eqnarray}

\noindent In the above equation, the dot over a symbol is the differentiation with respect to an affine parameter. From this Lagrangian, we get the generalized momenta,

\begin{eqnarray}
p_t &=& -f(r)\dot{t}=-E,\nonumber\\
p_r &=& \frac{\dot{r}}{f(r)},\nonumber\\
p_{\phi} &=& r^2\dot{\phi}=L.
\label{eq3.2}
\end{eqnarray}

\noindent Due to the Killing vector fields $\frac{\partial}{\partial t}$ and $\frac{\partial}{\partial \phi}$ in the spacetime, $E$ and $L$ are constants and denote the energy and orbital angular momentum of the photon, respectively. Using Eqs. (\ref{eq3.1}), (\ref{eq3.2}) and the Hamiltonian for this system \cite{WL}

\begin{eqnarray}
\mathcal{H} &=& 2\left(p_{\mu}\dot{x}^{\mu}- \mathcal{L} \right)= -E\dot{t} +\frac{\dot{r}^2}{f(r)} +L\dot{\phi}=0,
\label{eq3.3}
\end{eqnarray}

\noindent we obtain the equation of radial motion

\begin{eqnarray}
\dot{r}^2 +V(r)=0,
\label{eq3.4}
\end{eqnarray}

\noindent where $V(r)=- E^2+ \frac{L^2}{r^2}f(r)$ is an effective potential and determines the position of an unstable orbit. This position satisfies the condition

\begin{eqnarray}
V(r)=0, \quad\quad \frac{\partial V(r)}{\partial r} = 0, \quad\quad \frac{\partial^2 V(r)}{\partial r^2} <0.
\label{eq3.5}
\end{eqnarray}

\noindent Solving the middle equation of the above equations yields four roots. But only a root corresponds to this position, and this root is the radius of the photon sphere. Using the first and third equations, we can exclude the other three roots and get the radius $r_{ps}$. Due to the complexity of the expression $r_{ps}$, we do not write it here. Its specific values are calculated and listed in Table 1. As described in \cite{ZZLG,ZZLG2,ZZLG3}, $\alpha$ cannot be too negative, because the metric function may not be real inside the event horizon when $\alpha$ is too negative. The observational constraints on the coupling parameter $\alpha$ was found in \cite{CCFM}.

\begin{table}[htbp]
\begin{center}
\begin{tabular}
{|c|c|c|c|}
\hline
$Q$&
$\alpha $&
$r_{ps} $&
$ R_{Sh} $ \\
\hline
\raisebox{-3.00ex}[0cm][0cm]{0.1}&
-3.0&
3.889873182&
6.003984441 \\
\cline{2-4}
 &
-1.0&
3.367024270&
5.519636364 \\
\cline{2-4}
 &
0.8&
2.536680250&
4.811597083 \\
\hline
\raisebox{-3.00ex}[0cm][0cm]{0.4}&
-3.0&
3.826945093&
5.910587761 \\
\cline{2-4}
 &
-1.0&
3.286378595&
5.406970581 \\
\cline{2-4}
 &
0.8&
2.364595705&
4.628872985 \\
\hline
\raisebox{-3.00ex}[0cm][0cm]{0.7}&
-3.0&
3.682145249&
5.694974891 \\
\cline{2-4}
 &
-1.0&
3.095251850&
5.140325436 \\
\cline{2-4}
 &
0.5&
2.225312089&
4.397492519 \\
\hline
\raisebox{-3.00ex}[0cm][0cm]{1.0}&
-2.0&
3.141336116&
5.036599384 \\
\cline{2-4}
 &
-1.0&
2.746568247&
4.655643115 \\
\cline{2-4}
 &
-0.1&
2.129360012&
4.101527592 \\
\hline
\end{tabular}
\end{center}
\label{tab1}
Table 1 The radii of the photon sphere and shadow under the different values of Q and $\alpha$.
\end{table}

From Table 1, we find that the radius of the photon sphere is closely related to the charge and Gauss-Bonnet coupling constant. When the charge is fixed, the radius decreases with the increase of the value of the Gauss-Bonnet coupling constant. When the Gauss-Bonnet coupling constant is fixed, the radius increases with the decrease of the value of the charge.

A black hole shadow is an observable quantity. Its radius is calculated strictly via celestial coordinates. Through the calculation, it was found that the radius is related to the photon sphere. For a spherically symmetric black hole, the shadow radius is \cite{JAAM1}

\begin{eqnarray}
R_{Sh}= \left.\frac{r}{\sqrt{f(r)}}\right|_{r=r_{ps}}.
\label{eq3.6}
\end{eqnarray}

\noindent Inserting the radius of the photon sphere into the above equation, we get the shadow radius of the charged Einstein-Gauss-Bonnet black hole. The specific expression of the shadow radius is also very complex and is not presented here. The values of the radius are obtained by using the different values of $Q$ and $\alpha$. They are listed in Table 1.

It is found from the table that the shadow radius decreases with the increase of the value of the Gauss-Bonnet coupling constant when the charge is fixed and decreases with the increase of the value of the charge when the Gauss-Bonnet coupling constant is fixed. The radii of the photon sphere and shadow increase or decrease at the same time.

\section{Quasinormal modes, shadow and the correspondence with a test field}

As an important feature of black holes, QNMs are one of important ways to detect black holes. A lot of work has been done on QNMs by analytical and numerical approaches. The WKB approximation approach, which was first put forward in \cite{SW}, is an effective method to derive QNMs. This approach was extended to the $3rd$ order WKB approximation in \cite{IW} and higher order WKB approximation in \cite{IW1,IW2}, respectively. The $13th$ order WKB approximation approach was developed in \cite{IW2}. In this section, we adopt the $6th$ order WKB approximation approach \cite{IW1} to derive the QNMs by a scalar field perturbation in the four-dimensional charged Einstein-Gauss-Bonnet black hole. Although the accuracy of the QNMs derived by the $13th$ order WKB approximation approach is higher than that derived by the $6th$ order WKB approximation approach, the latter is enough for us to elaborate the correspondence between the QNMs in the eikonal limit and test field in the Einstein-Gauss-Bonnet-Maxwell gravity.

Let us start with a massless scalar field perturbation in the background of the metric (\ref{eq2.2}). The scalar field in the curved spacetime obeys

\begin{eqnarray}
\frac{1}{\sqrt{-g}}\partial_{\mu}\left(\sqrt{-g}g^{\mu\nu}\partial_{\nu}\Psi\right)=0,
\label{eq4.1}
\end{eqnarray}

\noindent where $\Psi$ is a function of coordinates $(t, r, \theta, \phi)$ and represents the scalar field. We adopt the following ansatz

\begin{eqnarray}
\Psi=\sum\limits_{l,m}{e^{-i\omega t}\Phi(r)r^{-1}Y_{lm}(\theta,\phi)}.
\label{eq4.2}
\end{eqnarray}

\noindent In the above equation, $\omega$ is the perturbation frequency, $e^{-i\omega t}$ denotes the time evolution of the scalar field, and $Y_{lm}(\theta, \phi)$ is a spherical harmonics. Inserting Eqs. (\ref{eq2.2}) and (\ref{eq4.2}) into Eq. (\ref{eq4.1}) and introducing a "tortoise" coordinate $dr_{\star}=\frac{dr}{f(r)}$ yield the Regge-Wheeler equation,

\begin{eqnarray}
\left[\partial_{r_{\star}}^2+ \omega^2 - \mathcal{V}(r)\right]\Phi(r)=0,
\label{eq4.3}
\end{eqnarray}

\noindent where

\begin{eqnarray}
\mathcal{V}(r)=f(r)\left[\frac{f^{\prime}(r)}{r}+\frac{l(l+1)}{r^2} \right].
\label{eq4.4}
\end{eqnarray}

\noindent The QNMs arise from the solution of the above differential equation under the specific boundary condition. The boundary condition is $r_{\star} \to \pm \infty$ which map the event horizon and infinity. The solution takes form $\Phi(r)\sim exp[-i\omega(t\mp)r_{\star}]$ with the oscillation modes $\omega=\omega_{R}-i\omega_{I}$. Our interest is focused on the fundamental modes where $n=0$. Since they have the least damping among the detected ringing signals and dominate the waveform of gravitational waves. Using the WKB approximation approach, we get the modes and list them in Table 2.1-Table 5.3. Here we did not calculate the modes with $l = n = 0$.  The reason is that this approach does not give a satisfactory accuracy for $n \ge l$. Nevertheless, it does not mean that this fundamental modes can not be calculated. One can get the modes by other approaches.

\begin{table}[htbp]
\begin{center}
\begin{tabular}{|c|c|c|c|c|}
\hline
&\multicolumn{2}{|c|} {QNMs}&
\multicolumn{2}{|c|} {Relative deviations}\\
\hline
&$\omega(WKB)$ &
$\omega (Sh)$&
Re $\omega$&
Im $\omega$ \\
\hline
l$=$1 &
0.232962-0.396877i &
0.249834-0.118605i&
-7.242{\%}&
70.115{\%} \\
\hline
l$=$2 &
0.443244-0.121925i &
0.416390-0.118605i&
6.059{\%}&
2.723{\%} \\
\hline
l$=$3 &
0.587432-0.121927i &
0.582946-0.118605i&
0.764{\%}&
2.725{\%} \\
\hline
l$=$4 &
0.751461-0.120694i &
0.749502-0.118605i&
0.261{\%}&
1.731{\%} \\
\hline
l$=$5&
0.917271-0.119989i &
0.916058-0.118605i &
0.132{\%}&
1.153{\%} \\
\hline
l$=$10 &
1.749280-0.118967i &
1.748839-0.118605i&
0.025{\%}&
0.304{\%} \\
\hline
l$=$50 &
8.411160-0.118620i&
8.411081-0.118605i&
0.0009{\%}&
0.013{\%} \\
\hline
\end{tabular}
\label{tab2.1}
\end{center}
Table 2.1 The QNMs derived by the WKB approximation approach and shadow radius and the relative deviations. In the calculation, we let $Q=0.1$ and $\alpha=-3$.
\end{table}

\begin{table}[htbp]
\begin{center}
\begin{tabular}{|c|c|c|c|c|}
\hline
&\multicolumn{2}{|c|} {QNMs}&
\multicolumn{2}{|c|} {Relative deviations}\\
\hline
&$\omega (WKB)$ &
$\omega (Sh)$&
Re $\omega$&
Im $\omega$ \\
\hline
l$=$1&
0.259512-0.120429i&
0.271757-0.106983i&
-4.718{\%}&
11.165{\%} \\
\hline
l$=$2&
0.453456-0.109868i&
0.452928-0.106983i&
0.116{\%}&
2.626{\%} \\
\hline
l$=$3&
0.635009-0.108202i&
0.634100-0.106983i&
0.143{\%}&
1.127{\%} \\
\hline
l$=$4&
0.816074-0.107674i&
0.815271-0.106983i&
0.098{\%}&
0.642{\%} \\
\hline
l$=$5&
0.997130-0.107433i&
0.996442-0.106983i&
0.069{\%}&
0.419{\%} \\
\hline
l$=$10&
1.902680-0.107102i&
1.902299-0.106983i&
0.020{\%}&
0.111{\%} \\
\hline
l$=$50&
9.149230-0.106988i&
9.149153-0.106983i&
0.0008{\%}&
0.005{\%} \\
\hline
\end{tabular}
\label{tab2.2}
\end{center}
Table 2.2 The QNMs derived by the WKB approximation approach and shadow radius and the relative deviations. In the calculation, we let $Q=0.1$ and $\alpha=-1$.
\end{table}

\begin{table}[htbp]
\begin{center}
\begin{tabular}{|c|c|c|c|c|}
\hline
&\multicolumn{2}{|c|} {QNMs}&
\multicolumn{2}{|c|} {Relative deviations}\\
\hline
&$\omega (WKB)$ &
$\omega (Sh)$&
Re $\omega$&
Im $\omega$ \\
\hline
l$=$1&
0.316086-0.080425i&
0.311747-0.078850i&
1.373{\%}&
1.958{\%} \\
\hline
l$=$2&
0.522310-0.079271i&
0.519578-0.078850i&
0.523{\%}&
0.531{\%} \\
\hline
l$=$3&
0.729380-0.079047i&
0.727409-0.078850i&
0.270{\%}&
0.249{\%} \\
\hline
l$=$4&
0.936778-0.078967i&
0.935240-0.078850i&
0.164{\%}&
0.148{\%} \\
\hline
l$=$5&
1.144330-0.078928i&
1.143072-0.078850i&
0.110{\%}&
0.099{\%} \\
\hline
l$=$10&
2.182890-0.078872i&
2.182228-0.078850i&
0.030{\%}&
0.027{\%} \\
\hline
l$=$50&
10.49560-0.078851i&
10.495476-0.07885i&
0.001{\%}&
0.001{\%} \\
\hline
\end{tabular}
\label{tab2.3}
\end{center}
Table 2.3 The QNMs derived by the WKB approximation approach and shadow radius and the relative deviations. In the calculation, we let $Q=0.1$ and $\alpha=0.8$.
\end{table}

\begin{table}[htbp]
\begin{center}
\begin{tabular}{|c|c|c|c|c|}
\hline
&\multicolumn{2}{|c|} {QNMs}&
\multicolumn{2}{|c|} {Relative deviations}\\
\hline
&$\omega (WKB)$ &
$\omega (Sh)$&
Re $\omega$&
Im $\omega$ \\
\hline
l$=$1&
0.262187-0.414285i&
0.253782-0.121017i&
3.206{\%}&
70.789{\%} \\
\hline
l$=$2&
0.453712-0.122533i&
0.422970-0.121017i&
6.775{\%}&
1.237{\%} \\
\hline
l$=$3&
0.597166-0.124275i&
0.592158-0.121017i&
0.839{\%}&
2.622{\%} \\
\hline
l$=$4&
0.763557-0.123146i&
0.761346-0.121017i&
0.290{\%}&
1.729{\%} \\
\hline
l$=$5&
0.931902-0.122444i&
0.930534-0.121017i&
0.147{\%}&
1.165{\%} \\
\hline
l$=$10&
1.776960-0.121395i&
1.776473-0.121017i&
0.027{\%}&
0.311{\%} \\
\hline
l$=$50&
8.544080-0.121033i&
8.543990-0.121017i&
0.001{\%}&
0.013{\%} \\
\hline
\end{tabular}
\label{tab3.1}
\end{center}
Table 3.1 The QNMs derived by the WKB approximation approach and shadow radius and the relative deviations. In the calculation, we let $Q=0.4$ and $\alpha=-3$.
\end{table}

\begin{table}[htbp]
\begin{center}
\begin{tabular}{|c|c|c|c|c|}
\hline
&\multicolumn{2}{|c|} {QNMs}&
\multicolumn{2}{|c|} {Relative deviations}\\
\hline
&$\omega (WKB)$ &
$\omega (Sh)$&
Re $\omega$&
Im $\omega$ \\
\hline
l$=$1&
0.263484-0.124484i&
0.277420-0.108859i&
-5.289{\%}&
12.552{\%} \\
\hline
l$=$2&
0.463000-0.111939i&
0.462366-0.108859i&
0.137{\%}&
2.752{\%} \\
\hline
l$=$3&
0.648293-0.110147i&
0.647313-0.108859i&
0.151{\%}&
1.169{\%} \\
\hline
l$=$4&
0.833116-0.109586i&
0.832259-0.108859i&
0.103{\%}&
0.663{\%} \\
\hline
l$=$5&
1.017940-0.109331i&
1.017205-0.108859i&
0.072{\%}&
0.432{\%} \\
\hline
l$=$10&
1.942340-0.108983i&
1.941938-0.108859i&
0.021{\%}&
0.114{\%} \\
\hline
l$=$50&
9.339880-0.108864i&
9.339796-0.108859i&
0.0009{\%}&
0.005{\%} \\
\hline
\end{tabular}
\label{tab3.2}
\end{center}
Table 3.2 The QNMs derived by the WKB approximation approach and shadow radius and the relative deviations. In the calculation, we let $Q=0.4$ and $\alpha=-1.$
\end{table}

\begin{table}[htbp]
\begin{center}
\begin{tabular}{|c|c|c|c|c|}
\hline
&\multicolumn{2}{|c|} {QNMs}&
\multicolumn{2}{|c|} {Relative deviations}\\
\hline
&$\omega (WKB)$ &
$\omega (Sh)$&
Re $\omega$&
Im $\omega$ \\
\hline
l$=$1&
0.327346-0.0765758i&
0.324053-0.074864i&
1.006{\%}&
2.235{\%} \\
\hline
l$=$2&
0.542113-0.0753940i&
0.540088-0.074864i&
0.374{\%}&
0.703{\%} \\
\hline
l$=$3&
0.757579-0.0751335i&
0.756124-0.074864i&
0.192{\%}&
0.359{\%} \\
\hline
l$=$4&
0.973297-0.0750294i&
0.972159-0.074864i&
0.117{\%}&
0.220{\%} \\
\hline
l$=$5&
1.189130-0.0749761i&
1.188194-0.074864i&
0.079{\%}&
0.150{\%} \\
\hline
l$=$10&
2.268860-0.0748956i&
2.268371-0.074864i&
0.022{\%}&
0.042{\%} \\
\hline
l$=$50&
10.90990-0.0748655i&
10.909783-0.074864i&
0.001{\%}&
0.002{\%} \\
\hline
\end{tabular}
\label{tab1}
\end{center}
Table 3.3 The QNMs derived by the WKB approximation approach and shadow radius and the relative deviations. In the calculation, we let $Q=0.4$ and $\alpha=0.8.$
\end{table}

\begin{table}[htbp]
\begin{center}
\begin{tabular}{|c|c|c|c|c|}
\hline
&\multicolumn{2}{|c|} {QNMs}&
\multicolumn{2}{|c|} {Relative deviations}\\
\hline
&$\omega (WKB)$ &
$\omega (Sh)$&
Re $\omega$&
Im $\omega$ \\
\hline
l$=$1&
0.380439-0.437918i&
0.263390-0.127070i&
30.767{\%}&
70.983{\%} \\
\hline
l$=$2&
0.476503-0.122625i&
0.438983-0.127070i&
7.874{\%}&
-3.625{\%} \\
\hline
l$=$3&
0.620628-0.129982i&
0.614577-0.127070i&
0.975{\%}&
2.240{\%} \\
\hline
l$=$4&
0.793031-0.129229i&
0.790170-0.127070i&
0.361{\%}&
1.671{\%} \\
\hline
l$=$5 &
0.967561-0.128571i&
0.965764-0.127070i&
0.186{\%}&
1.167{\%} \\
\hline
l$=$10&
1.844370-0.127484i&
1.843731-0.127070i&
0.035{\%}&
0.325{\%} \\
\hline
l$=$50&
8.867580-0.127087i&
8.867467-0.127070i&
0.001{\%}&
0.013{\%} \\
\hline
\end{tabular}
\label{tab1}
\end{center}
Table 4.1 The QNMs derived by the WKB approximation approach and shadow radius and the relative deviations. In the calculation, we let $Q=0.7$ and $\alpha=-3.$
\end{table}

\begin{table}[htbp]
\begin{center}
\begin{tabular}{|c|c|c|c|c|}
\hline
&\multicolumn{2}{|c|} {QNMs}&
\multicolumn{2}{|c|} {Relative deviations}\\
\hline
&$\omega (WKB)$ &
$\omega (Sh)$&
Re $\omega$&
Im $\omega$ \\
\hline
l$=$1&
0.273221-0.136267i&
0.291810-0.113586i&
-6.804{\%}&
16.645{\%} \\
\hline
l$=$2&
0.487400-0.117201i&
0.486351-0.113586i&
0.205{\%}&
3.084{\%} \\
\hline
l$=$3&
0.682100-0.115066i&
0.680891-0.113586i&
0.176{\%}&
1.286{\%} \\
\hline
l$=$4&
0.876450-0.114416i&
0.875431-0.113586i&
0.116{\%}&
0.725{\%} \\
\hline
l$=$5&
1.070830-0.114122i&
1.069971-0.113586i&
0.080{\%}&
0.470{\%} \\
\hline
l$=$10&
2.043140-0.113726i&
2.042672-0.113586i&
0.023{\%}&
0.123{\%} \\
\hline
l$=$50&
9.824380-0.113592i&
9.824281-0.113586i&
0.001{\%}&
0.005{\%} \\
\hline
\end{tabular}
\label{tab4.2}
\end{center}
Table 4.2 The QNMs derived by the WKB approximation approach and shadow radius and the relative deviations. In the calculation, we let $Q=0.7$ and $\alpha=-1$.
\end{table}

\begin{table}[htbp]
\begin{center}
\begin{tabular}{|c|c|c|c|c|}
\hline
&\multicolumn{2}{|c|} {QNMs}&
\multicolumn{2}{|c|} {Relative deviations}\\
\hline
&$\omega (WKB)$ &
$\omega (Sh)$&
Re $\omega$&
Im $\omega$ \\
\hline
l$=$1&
0.343629-0.0803095i&
0.341103-0.078513i&
0.735{\%}&
2.237{\%} \\
\hline
l$=$2&
0.570174-0.0790292i&
0.568506-0.078513i&
0.293{\%}&
0.653{\%} \\
\hline
l$=$3&
0.797117-0.0787623i&
0.795908-0.078513i&
0.152{\%}&
0.317{\%} \\
\hline
l$=$4&
1.024250-0.0786622i&
1.023310-0.078513i&
0.092{\%}&
0.190{\%} \\
\hline
l$=$5&
1.251490-0.0786127i&
1.250713-0.078513i&
0.062{\%}&
0.127{\%} \\
\hline
l$=$10&
2.388130-0.0785405i&
2.387724-0.078513i&
0.017{\%}&
0.035{\%} \\
\hline
l$=$50&
11.483900-0.0785143i&
11.483817-0.078513i&
0.0007{\%}&
0.002{\%} \\
\hline
\end{tabular}
\label{tab4.3}
\end{center}
Table 4.3 The QNMs derived by the WKB approximation approach and shadow radius and the relative deviations. In the calculation, we let $Q=0.7$ and $\alpha=0.5.$
\end{table}

\begin{table}[htbp]
\begin{center}
\begin{tabular}{|c|c|c|c|c|}
\hline
&\multicolumn{2}{|c|} {QNMs}&
\multicolumn{2}{|c|} {Relative deviations}\\
\hline
&$\omega (WKB)$ &
$\omega (Sh)$&
Re $\omega$&
Im $\omega$ \\
\hline
l$=$1&
0.340785-0.358543i&
0.297820-0.133492i&
12.608{\%}&
62.768{\%} \\
\hline
l$=$2&
0.521267-0.136806i&
0.496367-0.133492i&
4.777{\%}&
2.422{\%} \\
\hline
l$=$3&
0.699962-0.136866i&
0.694913-0.133492i&
0.721{\%}&
2.465{\%} \\
\hline
l$=$4&
0.895992-0.135613i&
0.893460-0.133492i&
0.283{\%}&
1.564{\%} \\
\hline
l$=$5 &
1.093700-0.134901i&
1.092007-0.133492i&
0.155{\%}&
1.044{\%} \\
\hline
l$=$10&
2.085430-0.133863i&
2.084740-0.133492i&
0.033{\%}&
0.277{\%} \\
\hline
l$=$50&
10.026700-0.133507i&
10.026606-0.133492i&
0.0009{\%}&
0.011{\%} \\
\hline
\end{tabular}
\label{tab5.1}
\end{center}
Table 5.1 The QNMs derived by the WKB approximation approach and shadow radius and the relative deviations. In the calculation, we let $Q=1$ and $\alpha=-2.$
\end{table}

\begin{table}[htbp]
\begin{center}
\begin{tabular}{|c|c|c|c|c|}
\hline
&\multicolumn{2}{|c|} {QNMs}&
\multicolumn{2}{|c|} {Relative deviations}\\
\hline
&$\omega (WKB)$ &
$\omega (Sh)$&
Re $\omega$&
Im $\omega$ \\
\hline
l$=$1&
0.297584-0.168863i&
0.322190-0.123414i&
8.564{\%}&
26.914{\%} \\
\hline
l$=$2&
0.540091-0.128214i&
0.536983-0.123414i&
0.575{\%}&
3.744{\%} \\
\hline
l$=$3&
0.753793-0.125369i&
0.751776-0.123414i&
0.268{\%}&
1.559{\%} \\
\hline
l$=$4&
0.968111-0.124504i&
0.966569-0.123414i&
0.159{\%}&
0.875{\%} \\
\hline
l$=$5&
1.182620-0.124114i&
1.181362-0.123414i&
0.107{\%}&
0.564{\%} \\
\hline
l$=$10&
2.255990-0.123595i&
2.255328-0.123414i&
0.029{\%}&
0.146{\%} \\
\hline
l$=$50&
10.84720-0.123422i&
10.847051-0.123414i&
0.0009{\%}&
0.006{\%} \\
\hline
\end{tabular}
\label{tab5.2}
\end{center}
Table 5.2 The QNMs derived by the WKB approximation approach and shadow radius and the relative deviations. In the calculation, we let $Q=1$ and $\alpha=-1.$
\end{table}

\begin{table}[htbp]
\begin{center}
\begin{tabular}{|c|c|c|c|c|}
\hline
&\multicolumn{2}{|c|} {QNMs}&
\multicolumn{2}{|c|} {Relative deviations}\\
\hline
&$\omega (WKB)$ &
$\omega (Sh)$&
Re $\omega$&
Im $\omega$ \\
\hline
l$=$1&
0.371004-0.097845i&
0.365717-0.097172i&
1.425{\%}&
0.688{\%} \\
\hline
l$=$2&
0.612625-0.097468i&
0.609529-0.097172i&
0.505{\%}&
0.304{\%} \\
\hline
l$=$3&
0.855547-0.097328i&
0.853341-0.097172i&
0.258{\%}&
0.160{\%} \\
\hline
l$=$4&
1.098870-0.097267i&
1.097152-0.097172i&
0.157{\%}&
0.098{\%} \\
\hline
l$=$5&
1.342370-0.097236i&
1.340964-0.097172i&
0.107{\%}&
0.066{\%} \\
\hline
l$=$10&
2.560760-0.097189i&
2.560022-0.097172i&
0.029{\%}&
0.017{\%} \\
\hline
l$=$50&
12.3126-0.097172i&
12.312486-0.097172i&
0.0008{\%}&
0{\%} \\
\hline
\end{tabular}
\label{tab5.3}
\end{center}
Table 5.3 The QNMs derived by the WKB approximation approach and shadow radius and the relative deviations. In the calculation, we let $Q=1$ and $\alpha=-0.1.$
\end{table}

There is a close connection between null geodesics and QNMs. It was proven in \cite{CMBWZ} that QNMs in the eikonal limit can be gotten via properties of null geodesics. In the eikonal limit, the QNMs takes form

\begin{eqnarray}
\omega = \Omega L-i\left(n+\frac{1}{2}\right)\lambda ,
\label{eq4.5}
\end{eqnarray}

\noindent where $\Omega=\sqrt{\frac{f^{\prime}(r)}{2r}}$ is the angular velocity at the unstable null geodesic orbit, $L$ is the angular momentum, $n$ is the overtone number, and $\lambda=\left.\sqrt{\frac{f(r)[2f(r)-r^2f^{\prime\prime}(r)]}{2r^2}}\right|_{r=r_{ps}}$ is the Lyapunov exponent that determines the instability time scale of the orbit. The connections between QNMs and photon spheres and between photon spheres and shadow radii have been found. Therefore, it is natural to conjecture that whether there is a connection between QNMs and shadow radii. This conjecture was verified in \cite{JAAM1}. The relation between the QNMs in the eikonal limit and shadow radii of the spherically symmetrical black holes was found. From this relation, the QNMs can be rewritten as

\begin{eqnarray}
\omega = R_{Sh}^{-1}\left(l+\frac{D-3}{2}\right) -i\left(n+\frac{1}{2}\right)\lambda ,
\label{eq4.6}
\end{eqnarray}

\noindent where $D$ is the dimension of spacetime. In the eikonal limit, the term $\frac{D-3}{2}$ in Eq. (\ref{eq4.6}) is neglected. However, we do not neglect it, since the QNMs at the small multiple number is also in our discussion. Using Eq. (\ref{eq4.6}), we get the QNMs of the charged Einstein-Gauss-Bonnet black hole and list them in Table 2.1-Table 5.3. In the tables, $\omega (WKB)$ denotes the QNMs derived by the $6th$ order WKB approximation approach, and $\omega (Sh)$ denotes the QNMs gotten by Eq. (\ref{eq4.6}). The relative deviations of the real (imaginary) parts of the QNMs calculated by Eq. (\ref{eq4.6}) from those calculated by the WKB approximation approach are listed in the last two columns. From these tables, we find the following phenomena.

1. For the fixed charge and same multiple number, when the value of the Gauss-Bonnet coupling constant increase, the values of the real parts of the QNMs increase and those of the imaginary parts decrease. The coincidence degrees of the QNMs derived by the two approaches increase with the increase of the value of $\alpha$. For $\alpha = - 3$ and $l=1$, the degree of coincidence is the smallest. The reason may be that the value of $\alpha$ cannot be too negative.

2. For the fixed charge and Gauss-Bonnet coupling constant, the result derived by the WKB approximation approach shows that the values of the real parts increase and those of the imaginary parts decrease slowly when the value of the multiple number increases. However, when $l = 1$, the imaginary parts' values are the largest and the degrees of coincidence are the smallest. When $l = 3$, the values are the second largest. The reason may be related to the condition of the accuracy for the WKB approximation approach.

3. For the fixed Gauss-Bonnet coupling constant and multiple number, when the value of the charge increases, the values of the real parts and imaginary parts calculated by the WKB approximation approach increase at the same time, but the coincidence degrees of the QNMs obtained by the two approaches decrease. For example, when $\alpha = - 1$ and $l = 5$, we can find the QNMs and the degrees of coincidence for the different value of $Q$.

4. The values in the tables show that when the value of the multiple number is large, the QNMs gotten by the two approaches are consistent. In fact, when $l$ is small, the QNMs are in good agreement with each other except for the values at $\alpha = -3$. This result is full in consistent with those derived in \cite{GM,BKI}.

\section{Conclusions}

In this paper, we investigated the photon sphere, shadow and QNMs of the four-dimensional charged Einstein-Gauss-Bonnet black hole. The QNMs were gotten by the $6th$ order WKB approximation approach and shadow radius, respectively. When the value of the multiple number is large, the QNMs calculated by the two approaches are in good agreement, which shows the correspondence between the shadow and test field in the four-dimensional Einstein-Gauss-Bonnet-Maxwell gravity. Eq. (\ref{eq4.6}) is obtained under the condition of the large multiple numbers. However, we found that when the value of the multiple number is small, except for the values at $\alpha = -3$ and $l=1$, the degrees of coincidence of the QNMs obtained by the two approaches are also in good agreement. The degrees of coincidence increase with the increases of the value of the Gauss-Bonnet coupling constant. When the value of the Gauss-Bonnet coupling constant increase, the values of the real parts of the QNMs increase and those of the imaginary parts decrease. It shows that the Gauss-Bonnet coupling constant plays an important role in the QNMs. This investigation reveals a potential relationship between the black hole shadow and gravitational wave.

\vspace*{2.0ex}
\noindent \textbf{Acknowledgments}

\noindent This work is supported by NSFC (Grant Nos. 11711530645, 11205125), the Program for Innovative Youth Research Team in University of Hubei Province of China (Grant Nos. T201712) and FXHU (Z201021).

\end{document}